\documentclass[conference, 10pt]{IEEEtran}

\usepackage{epsfig, amssymb, amsmath, amsthm, subcaption,hyperref, setspace}
\theoremstyle{definition} \newtheorem{definition}{Definition}

\DeclareMathOperator*{\argmax}{arg\,max}
\DeclareMathOperator*{\argmin}{arg\,min}

\begin{document}

\title{Linking Correlated Network Flows through Packet Timing: a Game-Theoretic Approach}

\author{\IEEEauthorblockN{Juan A. Elices}
\IEEEauthorblockA{University of New Mexico\\
Email: jelices@ece.unm.edu}
\and
\IEEEauthorblockN{Fernando P\'erez-Gonz\'alez}
\IEEEauthorblockA{Universidad de Vigo\\
Email: fperez@gts.uvigo.es}
}

\maketitle

\begin{abstract}
Deciding that two network flows are essentially the same is an important problem in intrusion detection or in tracing anonymous connections. A stepping stone or an anonymity network may try to prevent flow correlation by delaying the packets, introducing chaff traffic, or even splitting the flow in several subflows. 

We introduce a game-theoretic framework for this problem. The framework is used to derive the Nash equilibrium under two different adversary models: the first one, when the adversary is limited to delaying packets, and the second, when the adversary also adds dummy packets and removes packets from the flow. As the optimal decoder is not computationally feasible, we restrict the possible decoder to one that estimates and compensates the attack. Our analysis can be used for understanding the limits of flow correlation based on packet timings under an active attacker.
\end{abstract}

\IEEEpeerreviewmaketitle

\section{Introduction}
Network attackers intentionally hide their identity to avoid prosecution.  A broadly-used way of achieving this anonymity is relaying the traffic through a chain of compromised hosts called stepping stones~\cite{StHe:95}. Intrusion detection and tracing back an attack require deciding that two flows are correlated. Linking network flows can also be used to compromise low-latency anonymous networks, such as Tor.

Two approaches exist for finding correlated flows: passive analysis and active watermarks. They differ in whether the flow is modified or not. In general, an active watermark needs shorter sequences but at the expense of being detectable~\cite{LuZhZhPeLe:11}.

An adversary (AD), for instance a stepping stone or an anonymous network, may take countermeasures to prevent the correlation such as introducing delays to packets or adding dummy packets to the flow, or even more drastic measures, e.g. dividing the flow into different subflows each taking different paths. To the best of our knowledge, in previous works, only \cite{DoFlShPaCoStL:02} and \cite{BlSoVe:04} consider an AD, and in both the AD is limited to delaying packets.

This paper studies the limits of traffic analysis, passive or active, in an adversarial environment. The most natural solution to avoid the loop of proposing an attack and creating an \emph{ad-hoc} solution is to cast the problem into a game-theoretic framework and look for the optimum strategies that the players, traffic analyst (TA) and AD, should adopt.  A game-theoretic framework similar to the proposed one has been used in other contexts such as Information Hiding~\cite{MoOs:03} or Source Identification~\cite{BaTo:13}.

The rest of the paper is organized as follows: in Section \ref{sec:not} we introduce the notation, together with some basic concepts of game theory. Section \ref{sec:mod} presents a rigorous deﬁnition of the traffic analysis game. Section \ref{sec:del} solves the problem under an AD model that only introduces delays. Section \ref{sec:aad} deals with a stronger AD problem that also adds chaff traffic and divides the flow. Conclusions are presented in Section \ref{sec:con}.

\section{Notation and Basic Concepts}\label{sec:not}
We use the following notation. Random variables are denoted by capital letters (e.g., $X$), and their individual realizations by lower case letters (e.g., $x$). The domains over which random variables are defined are denoted by script letters (e.g., $\mathcal{X}$). Sequences of $n$ random variables are denoted with $X^n$ if they have random nature or by $x^n$ if they are deterministic. $X_{i}$ or $x_{i}$ indicate the $i$−th element of $X^n$ or $x^n$, respectively. The  probability distribution function (pdf) of a random variable $X$ is denoted by $f_X(x),\, x\in \mathcal{X}$. We use the same notation to refer to pdf of sequences, i.e. $f_{X^n}(x^n), \, x^n \in \mathcal{X}^n$. When no confusion is possible, we drop the subscript in order to simplify the notation. We denote with $\Delta$ the difference operation of a sequence, i.e $\Delta x^n=\{x_2-x_1, \dots, x_n-x_{n-1}\}$ and with $\Delta x_i=x_{i+1}-x_i$ the $i$th element of this sequence.

\subsection{Game Theory}

Game theory is the mathematical study of interaction among independent, self-interested agents. Formally, a two-player game is defined as a quadruple $G(A_1,A_2, u_1, u_2)$, where  $A_i= \{a_{i,1},\dots a_{i,n_i}\}$ are the actions available to the $i$th player, and $u_i: A_1 \times A_2 \mapsto \mathbb{R}, \; i = 1,2$ is the utility function or payoff of the game for player $i$. An action profile is the double $a \in A_1 \times A_2$.
We  are interested in zero-sum games, where $u_1(a)+ u_2(a)=0, \forall a \in A_1 \times A_2$, which means that the gain (or loss) of utility of player 1 is exactly balanced by the losses (or gains) of the utility of player 2. In this case, we can simplify the game notation to a triplet $G(A_1,A_2, u)$, where $u=u_1=-u_2$. 

We say that an action profile $(a_{1,i^*}; a_{2,j^*})$ represents a Nash
equilibrium (NE)  if 
$u(a_{1,i^*}; a_{2,j^*}) \geq u(a_{1,i}; a_{2,j^*})\;\; \forall a_{1,i} \in A_1$ and 
$u(a_{1,i^*}; a_{2,j^*}) \leq u(a_{1,i^*}; a_{2,j})\;\; \forall a_{2,j} \in A_2$,
intuitively, this means that none of the players can improve its utility by modifying his 
strategy assuming the other player does not change his own.

Games can be classified in simultaneous games, where both players move unaware of the other player action, and sequential games, where later players have some knowledge about earlier actions. In sequential games an action profile is a subgame perfect equilibrium (SPE) if it represents a NE of every subgame of the original game. Therefore, a SPE  is a refinement of the NE that eliminates non-credible threats.

\section{Traffic Analysis Game}\label{sec:mod}

\begin{figure}
  \centering
    \includegraphics[width=0.7 \columnwidth]{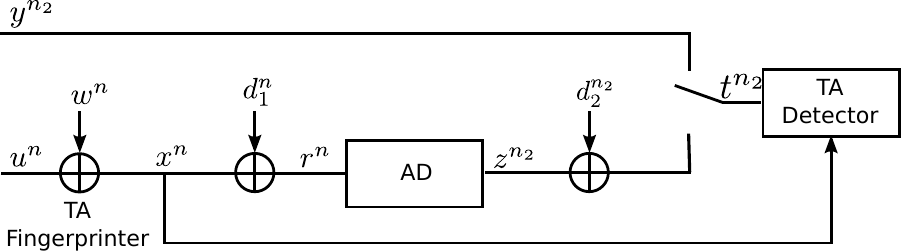}
  \caption{Model of the TAG game}
  \label{fig:tag}
\end{figure}

The Traffic Analysis game (TAG) is represented in Figure~\ref{fig:tag}. In this game, there are two players: TA and AD.
 
The task of the TA is to create a test to accept or reject the hypothesis that a flow $w^m$ is indeed the same flow as a known one, $x^n$, that can be the original flow (passive traffic analysis) or the output of a watermarker. In this paper we do not consider the problem of how to create the watermark. Furthermore, all the simulations are done in a passive analysis situation.

On the other side, the goal of the AD is to modify the flow in such a way that the TA decides that this sequence is not a modified version of $x^n$. In order to do this, AD can delay any packet at most $A_{max}$ seconds. We denote by $a^n$ the sequence of delays added by AD. In Section \ref{sec:aad}, we propose a more powerful attacker model that can add up to $P_A\cdot n$ dummy packets and remove $P_L\cdot n$ packets, assuming they are sent through a different path, hence $P_A$ is the maximum ratio between chaff traffic and original traffic and $P_L$ is the maximum probability of a packet being removed. 

As shown in Figure \ref{fig:tag} the flow suffers two additional delays $D_1^n$ and $D_2^m$ that are due to the network between where $x^n$ is measured (TA-1) and AD, and between AD and where $w^m$ is measured (TA-2), respectively. We represent by $D$ the delay that suffers a packet from TA-1 to TA-2, i.e. $D=D_1+D_2$. Note that $\Delta D$ is the packet delay variation (PDV), also called jitter.

Let $Y^m$ represent flows without any relation to $x^n$, but from the same application. We assume that $f_{\Delta Y}(\Delta y)$ is known by both players, and define the hypotheses: $H_0$: $w^m$ is not a modified version of $x^n$, and $H_1$: $w^m$ is a modified version of $x^n$.

We define the TAG game as follows:
\begin{definition} 
The $TAG(A_{TA};A_{AD}; u)$ is a simultaneous, zero-sum game played by the FA and the AD, where
\begin{itemize}
\item The set of actions the FA can choose from, i.e. $A_{TA}$, is the set of acceptance regions for $H_1$ for which the probability of false positive  (i.e., accepting $H_1$ when $H_0$ is true) is below a certain threshold $\eta$. Note that the acceptance region can be different for each sequence. Therefore, 
$A_{TA}=\{\Lambda_1(x^n): Pr(y^m \in \Lambda_1(x^n))<\eta\}$.
\item The set of possible attacks that the AD can choose from, i.e. $A_{AD}$. It depends on the assumed adversary model.
\item The utility function is the probability of detection, (i.e., accepting $H_1$ when $H_1$ is true), namely:
$u(\Lambda_1(x^n), A_{AD})= Pr( w^m|H_1 \in \Lambda_1(x^n))$.
\end{itemize}
\end{definition}

As the TA can choose its action $\Lambda_1(x^n)$ after knowing $w^n$, the solution to the game is
\begin{equation}\label{eq:mim}
u=\min\limits_{A_{AD}}\max\limits_{A_{TA}}u(A_{TA},A_{AD}).
\end{equation}

\section{Delaying Adversary model}\label{sec:del}
This section derives the detector under an AD that is limited to delaying the packets constrained to $A_{max}$ seconds. Under this condition, there exists a one-to-one correspondence between $x^n$ and $w^m$, consequently $m=n$.

We confine the detector to those based on first-order statistics of the inter-packet delay (IPD), i.e., $\Delta x$ and $\Delta w$, for feasibility reasons. Among those, the FA constructs the optimal detector which, according to Neyman-Pearson Lemma, is the likelihood ratio test
\begin{align} \label{eq:det1}
&\Lambda_{AD}(w^n,x^n,f_{\hat{A}^n|X^n})= \frac{\ell(H_1|\Delta w^n,\Delta x^n)}{\ell(H_0|\Delta w^n,\Delta x^n)}=\sum_{i=1}^{n-1} \log \left( \iint_{\hat{\mathcal{A}}^2}  \right. \nonumber \\ &  \frac{f_{\Delta D}(\Delta r_i -\Delta \hat{a}_i) f_{\hat{A}_{i,i+1}|X_{i,i+1}}(\hat{a}_i,\hat{a}_{i+1}|x_{i,i+1}) }{f_{\Delta Y}({\Delta w_i})}d\hat{a}_id\hat{a}_{i+1}\bigg),
\end{align}
where $r^n$ is $w^n-x^n$, $\ell$ represents the log-likelihood function, and $f_{\hat{A}^n}$ the estimated joint pdf for $A^n$. Hence, the test chooses $H_1$ when $\Lambda_{AD}(w^n,x^n,f_{\hat{A}^n|X^n})\geq \epsilon$ where $\epsilon$ is a threshold that we fix to achieve a certain probability of false positive.  In this case, the TA actions are all the possible joint pdfs for $\hat{A}^n$ given $x^n$, i.e. $A_{TA}=f_{\hat{A}^n|X^n}(a^n|x^n)$. Unfortunately, calculating the joint pdf that maximizes $\Lambda_{AD}$ is a computationally intractable problem. 

Therefore, we restrict the detector, at the expense of losing the optimality, to one that estimates the sequence $\hat{a}^n$ instead of its joint pdf. In this case, the likelihood ratio test is
\begin{equation} \label{eq:det2}
\Lambda_{AD}(w^n,x^n,\hat{a}^n)= \sum_{i=1}^{n-1} \log \left(\frac{f_{\Delta D}(\Delta_i ( r^n-\hat{a}^n))}{f_{\Delta Y}(\Delta_i w^n)} \right),
\end{equation} 
and the TAG game is modified as follows:
\begin{align}
A_{TA}&=\{\hat{a}^n: \forall i \in [0, n],\; 0\leq \hat{a}_i \leq A_{max}\ \}\\
A_{AD}&=\{a^n: \forall i \in [0, n],\; 0\leq a_i \leq A_{max}\}\\
u(\hat{a}^n, a^n)&=Pr(\Lambda_{AD}(x^n+a^n+D^n,x^n,\hat{a}^n)> \epsilon), \label{eq:u1}\\
 \text{where }&Pr(\Lambda_{AD}(Y^n,x^n,\hat{a}^n) \leq \epsilon) = \eta. \label{eq:u2}
\end{align}
From \eqref{eq:mim}, we obtain the solution to the game
\begin{align}
A_{TA}=&\argmax_{\hat{a}}\Lambda_{AD}(w^n,x^n,\hat{a}^n) \label{eq:ta1} \\
A_{AD}=&\argmin_{a^n} Pr(\max_{\hat{a}^n}\Lambda_{AD}(x^n+a^n+D^n,x^n,\hat{a}^n) > \epsilon)\label{eq:ad1}.
\end{align}

As the AD must decide its action in real time and \eqref{eq:ad1} is computationally expensive, we approximate it by
\begin{equation}
A_{AD} \approx \argmin_{a^n} \sum_{i=1}^{n-1} \log f_{\Delta Y}(\Delta x_i+ \Delta a_i)\label{eq:ad2}.
\end{equation} 
which is a good approximation when $\Delta D$ has zero-mean and its variance is much smaller than $\Delta Y$ (as it is the case in practice). Note that under this approximation AD is maximizing the likelihood of $x^n+a^n$ coming from $y^n$, i.e., making the sequence as typical as possible.

\subsection{Performance}

In this section we construct a simulator and present the scenarios we use in the remaining of the paper.  Afterwards, we compare the detector performance between the AD choosing its action optimally and an AD that selects its attack randomly.

\subsubsection{Simulator and Scenarios}

Simulations are carried out in the following way. First, we randomly generate one sequence, $X^n$. Then we generate $10^6$ sequences $Y^n$ and we calculate $\epsilon$
such that \eqref{eq:u2} holds. To deal with the probabilistic nature of the delay, we repeat the following 50 times: we apply $D_1^n$, next, we implement the adversary action according to \eqref{eq:ad2}. Subsequently, we introduce another delay $D_2^n$. Finally, we calculate the utility for this $X^n$ using \eqref{eq:u1}. We repeat the whole process for 1000 different $X^n$, the plots show the average utility for the $1000$ sequences, $\bar{u}$. Recall that the utility function is the probability of detection. In order to show that AD is acting rationally we compare it with an AD that introduces delays randomly, i.e., $a^n$ is an independent and identically distributed (i.i.d.) sequence uniformly distributed between 0 and $A_{max}$.

Note that \eqref{eq:det2} needs an estimation of $f_{\Delta D}$ and $f_{\Delta Y}$, to this end we apply kernel smoothing techniques~\cite{BoAz:97}.  As it is customary, we separate the data into two subsets: training, to obtain the pdfs, and test, used in the simulator, using 50\% of the samples for each.

Scenario A represents a stepping stone that forwards SSH traffic inside the Amazon Web Services network. TA-1, AD and TA-2 (cf. Figure \ref{fig:tag}) are EC2 instances located in Virginia, Oregon and California, respectively.  We use the IPDs from $8746$ replayed SSH connection captures with 64 million packets from \cite{KoHeAbYe:04} and \cite{BaSaPrVa:10}. The simulated delays correspond to Scenario 10 from \cite{ElPe:13}.

Scenario B simulates a web page accessed from Tor network whose real origin is to be found. In it, TA-1 corresponds to the web server, AD to the Tor entry relay, and TA-2 to the client. For instance, this case can correspond to a company in whose forum an anonymous insulting post has been placed using Tor and it is to be known whether the source comes from an employee within the company.  We use the IPDs of $113690$ replayed HTTP connections that sum around 139 million packets taken from the same repositories.
The delays correspond to the measurements of Scenario 11 from \cite{ElPe:13}. 

\begin{figure}
  \centering
  \begin{subfigure}[b]{\columnwidth}
	\centering
	\includegraphics[width=0.85\textwidth]{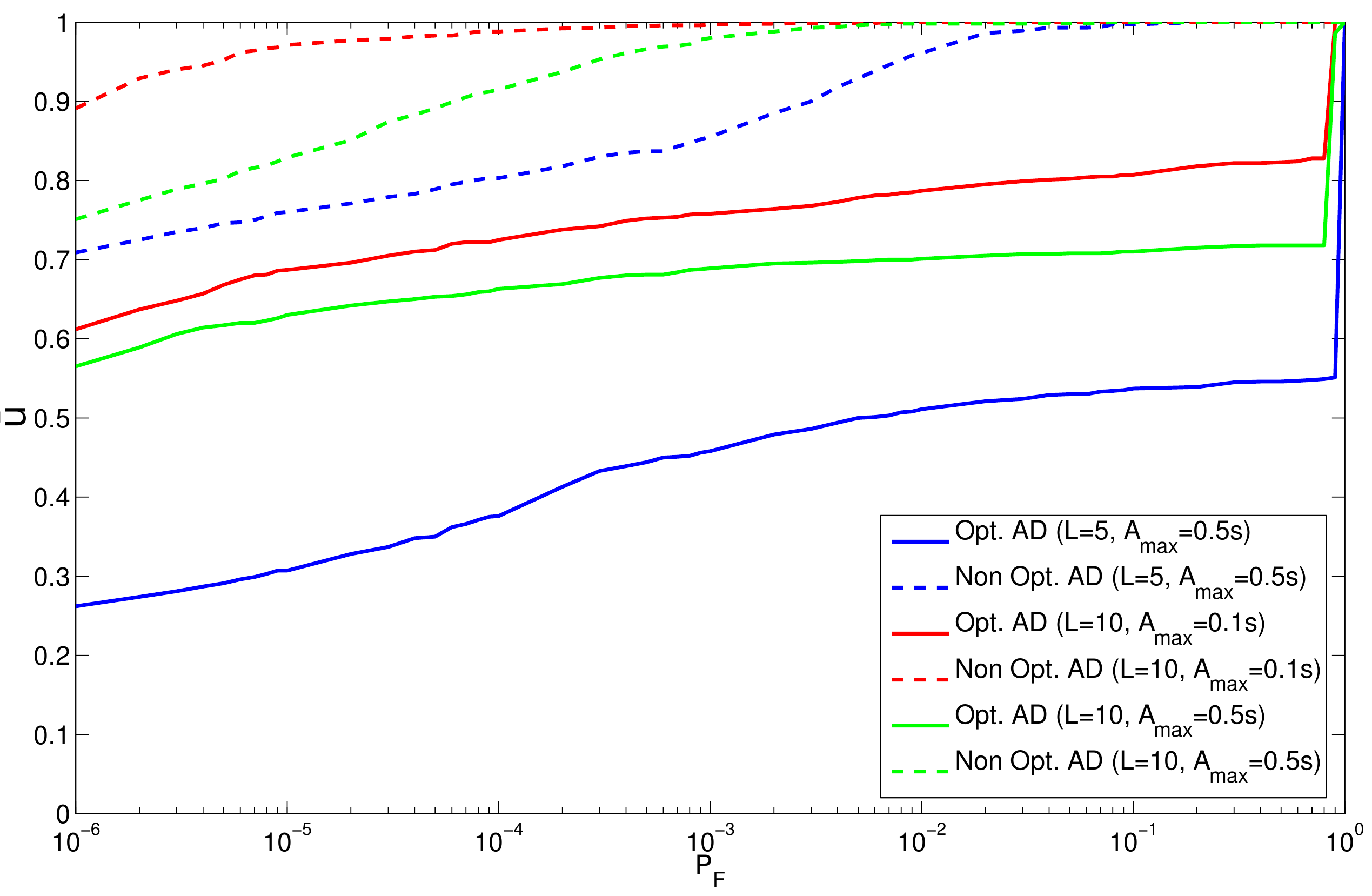}
	\caption{Scenario 1}
	\label{fig:Jitsc1}
  \end{subfigure}
  \begin{subfigure}[b]{\columnwidth}
	\centering
	\includegraphics[width=0.85\textwidth]{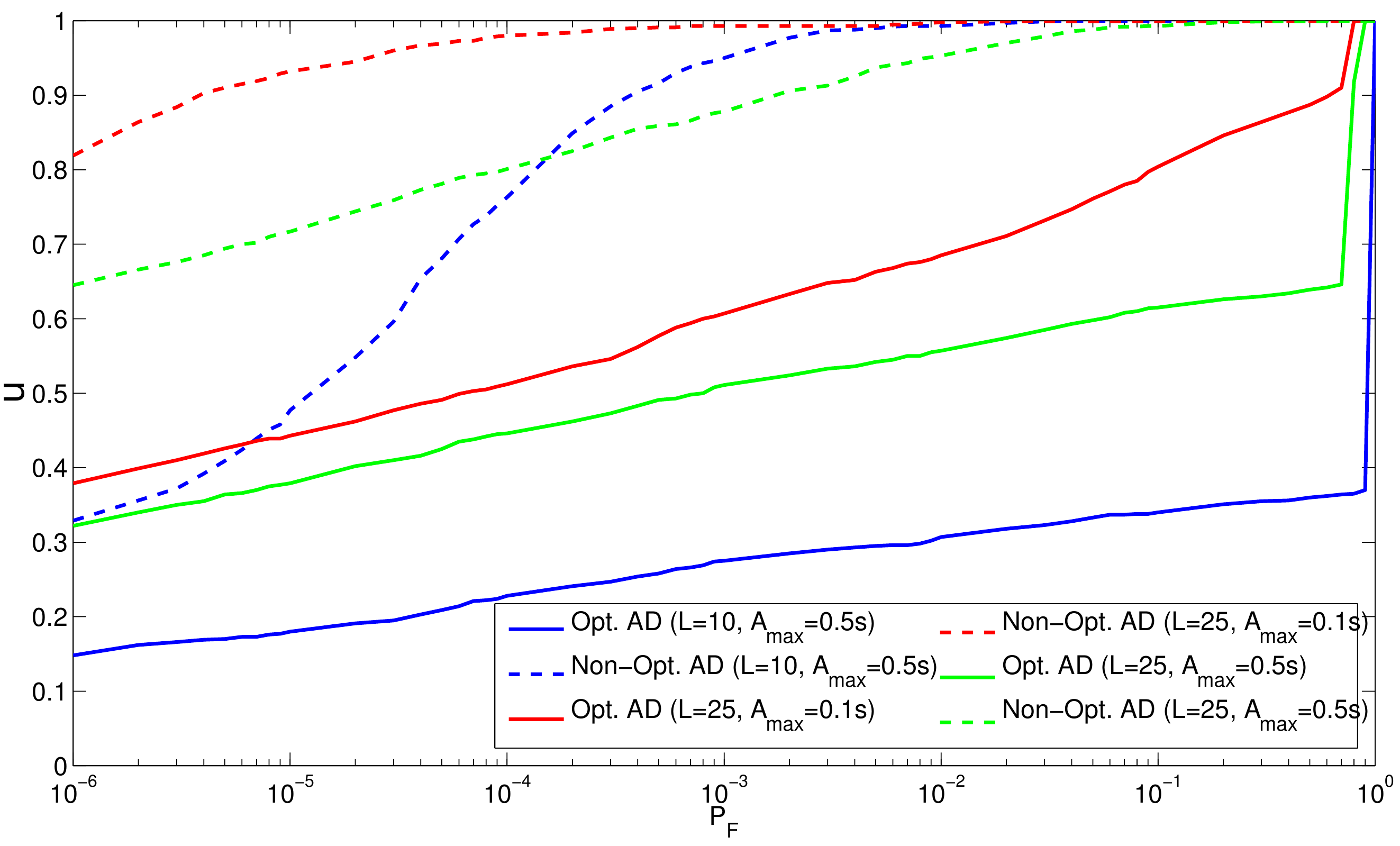}
	\caption{Scenario 2}
	\label{fig:Jitsc2} 
  \end{subfigure}
  \caption{Performance under a delaying AD model.}
\end{figure}

Results are depicted in Figures \ref{fig:Jitsc1} and \ref{fig:Jitsc2} for Scenario A and B, respectively. We see that if AD chooses the delays optimally, the impact of the attack is much larger than using random delays.

\section{Chaff traffic and flow splitting}\label{sec:aad}

Hitherto, we have assumed that there exists a one-to-one relation between the flows at the creator and detector; i.e., no packets are added or removed. In this section, we make a robust test to repacketization and study the game when, in addition to random delays, AD can also insert chaff traffic and divide the flow, and TA-2 observes only one subflow.

To deal with packet addition and removal, TA matches each packet from $x^n$ with the most likely from $w^m,\; m\leq(1+P_A)n$, denoting this as $i \rightarrow j$, as follows: 
$|x_i-(w_j-\rho)| < |x_i-(w_k-\rho)|,\; \forall k\neq j$, and to avoid considering it lost
$|x_i-(w_j-\rho)|>\gamma$, where $\rho$ is a synchronization constant obtained through an exhaustive search (auto-synchronization property), and $\gamma$ is a threshold
for which a packet is considered lost. To prevent that two packets $i_1$ and $i_2$ are matched to the same $j$ packet, the set of possible matching candidates for the later packet is reduced to the non-matched ones. Also $\gamma$  should be big enough so that the probability that a packet is considered lost when it actually is not is very small, for instance our simulation uses $\gamma=Q_{|\Delta D|,0.999}+A_{max}$ where $Q_{|\Delta D|,0.999}$ means the $0.999$-quantile of the absolute value of $\Delta D$. This procedure outputs two subsequences, $w^{n_2}$ and $x^{n_2}$, where $n_2\leq n$.

After this matching we apply a modification to $\Lambda_{AD}$ in \eqref{eq:det2} that takes into account the new sequence length as follows:
\begin{align} \label{eq:det3}
\Lambda_{AD}(w^{n_2},x^{n_2},\hat{a}^{n_2})&=(n-n_2)\log(P_L)+\sum_{i=1}^{n_2-1} \log \left(P_L\right. \nonumber\\ &\left.+(1-P_L)\frac{f_{\Delta D}(\Delta w_i-\Delta x_i-\Delta \hat{a}_i)} 
{f_{\Delta Y}(\Delta w_i)} \right),
\end{align}

We propose a new AD model that apart from adding delays as previously, it can remove up to $P_L \cdot n$ packets and add at most $P_A \cdot n$ dummy packets. Formally, the available actions for the adversary are:
\begin{align}
& A_{AD}=\{a^n \times l^n \times c^{n_A}: 0\leq a_i \leq A_{max}, l_i=\{0,1\} \nonumber \\ & \sum_{i=1}^n \frac{l_i}{n} \leq P_L,\; \frac{n_A}{n} \leq P_A,\,  m=n+n_A-\sum_{i=1}^n l_i\},
\label{eq:sad1}
\end{align}
where $l^n$ is a binary sequence that represents which packets are removed from the flow, and $c^{n_A}$ is the sequence with the timing of the $n_A$ introduced dummy packets. 

\subsection{Results}
To evaluate the proposed robust algorithm, we modify the simulator using the matching process as follows: longer sequences, i.e. $Y^{n+\lfloor P_An\rfloor}$, are needed to calculate $\epsilon$. Afterwards, AD chooses its action such that \eqref{eq:det3} is minimized. 
We compare the optimal AD with two non-optimal ADs: a) NO1 AD chooses its attack randomly, i.e $a^n$ is an i.i.d. sequence uniformly distributed between 0 and $A_{max}$, $l^n$ is a binary random sequence that contains $\lfloor nP_L  \rfloor$ `1's, and $c^{n_A}$ is an i.i.d. sequence uniformly distributed between the timing of the first packet and the timing of the last packet. The  b) NO2 AD selects $a^{n}$ to minimize \eqref{eq:det3} (optimally) but chooses $l^n$ and $c^{n_A}$ identically as the previous AD.  

\begin{figure}
  \centering
  \begin{subfigure}[b]{\columnwidth}
	\includegraphics[width=0.9\textwidth]{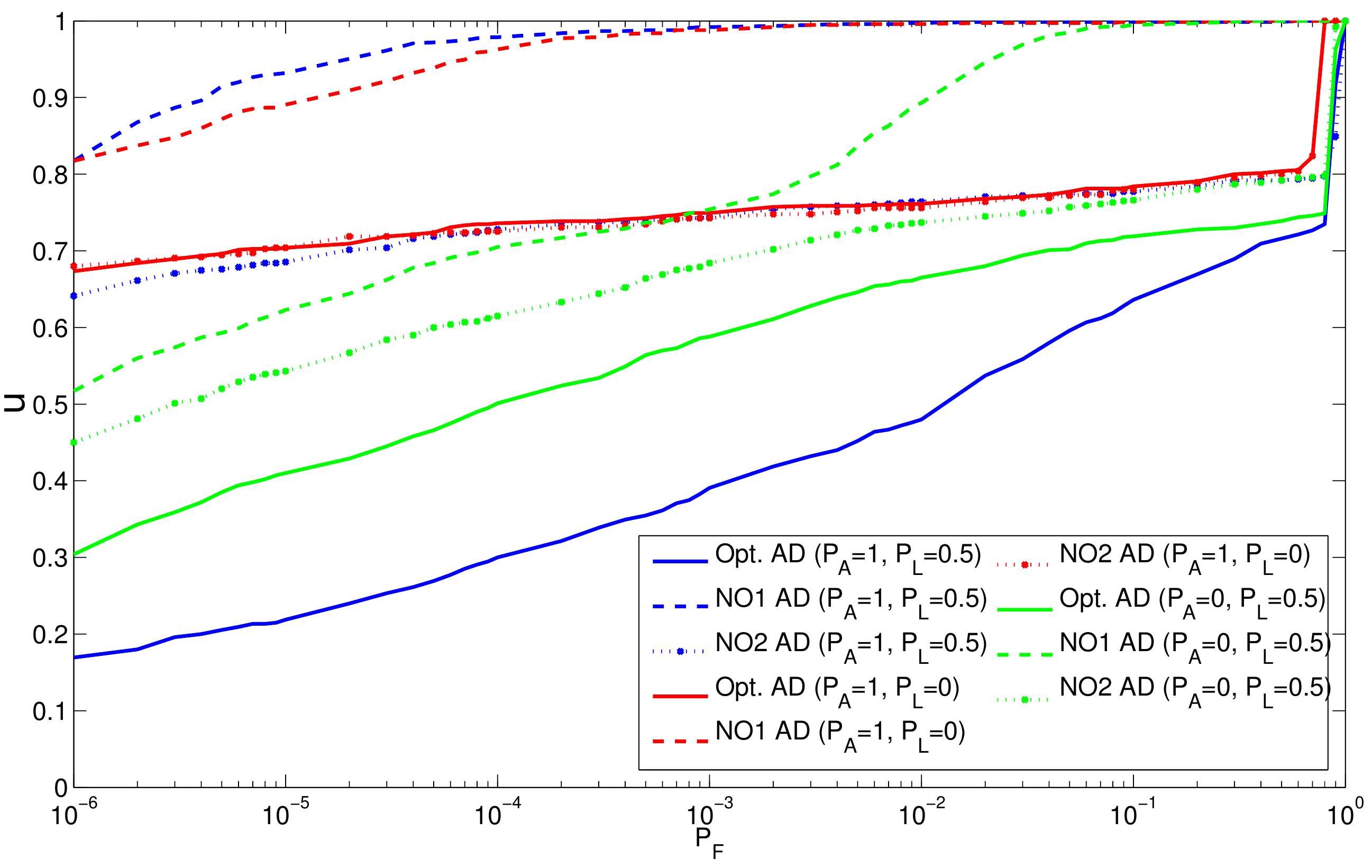}
    \caption{Scenario 1, L=20, $A_{max}=500$ms}
	\label{fig:AtSc1}
  \end{subfigure}
  \begin{subfigure}[b]{0.9\columnwidth}
	\includegraphics[width=\textwidth]{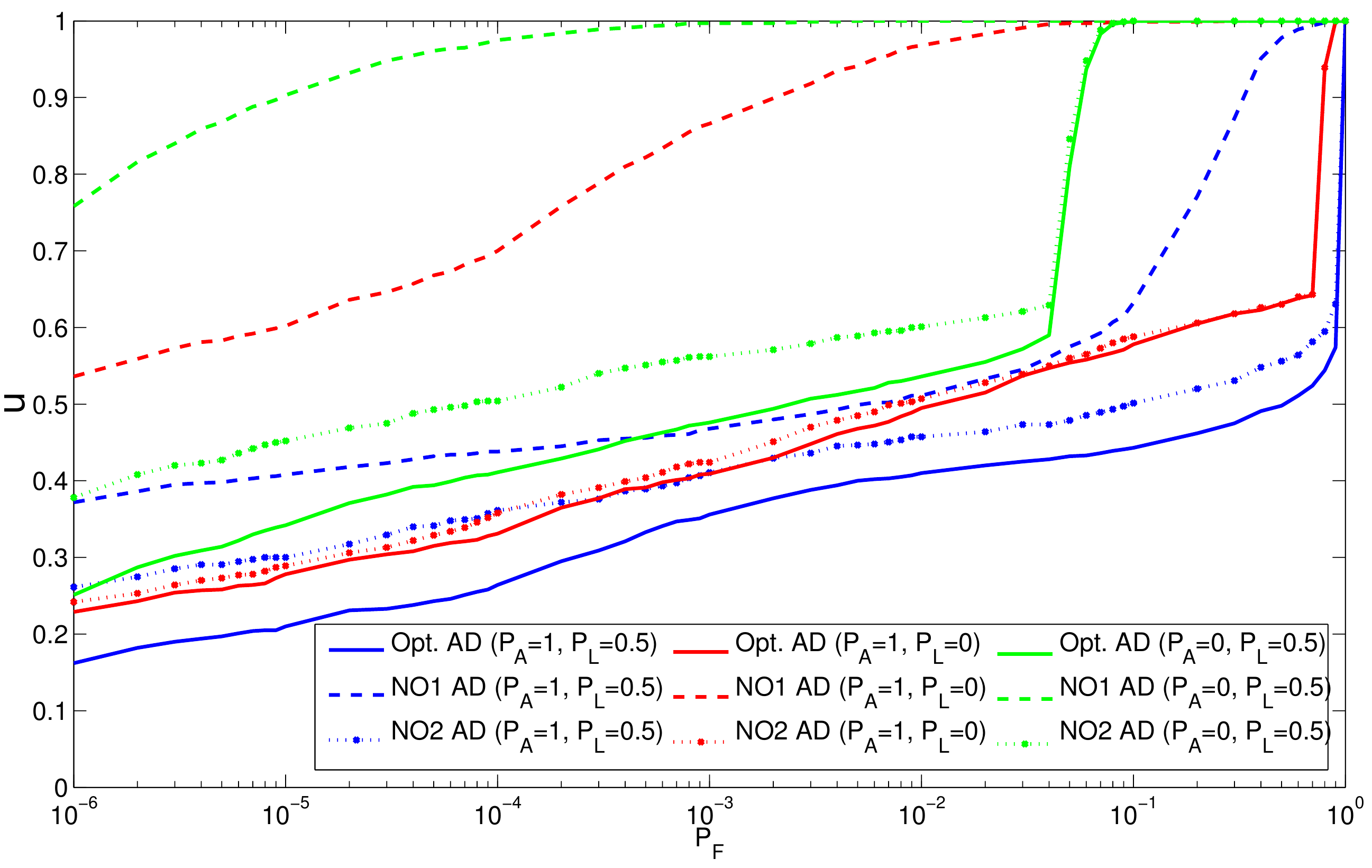}
	\caption{Scenario 2, L=30, $A_{max}=500$ms}
	\label{fig:AtSc2}
  \end{subfigure}
  \caption{Performance under chaff packets and removed packets.}
\end{figure}

Results are depicted in Figures \ref{fig:AtSc1} and \ref{fig:AtSc2}. We see that the chaff traffic location when there are no losses does not make a big impact (the Opt-AD and NO-2 AD lines difference is very small). The rest of the data confirm that AD can harm the TA considerably more when deciding its actions logically than acting randomly.

\section{Conclusions}\label{sec:con}
Using a game theoretic framework we have analyzed the flow linking problem. This problem consists in deciding if two flows are linked, having an adversary in the middle that tries to impair the correlation. Using this framework we obtain the optimal choices if both players act rationally. However, in a real implementation finding these optimal choices is not computationally feasible. This made us restrict the decoder to those that estimate the attack. We apply it to two different adversaries, one that only delays packets and another that can also add and drop packets from the flow. While in the present work the original flow is given, a future work will allow this flow to be modified through a watermark.

\section*{Acknowledgment}
Research supported by Iberdrola Foundation through the Prince of Asturias Endowed
Chair in Information Science and Related Technologies.

\begin{spacing}{0.9}
\bibliographystyle{IEEEtran}
\bibliography{../bib/watermark,../bib/gametheory,../bib/traces}

\begin{thebibliography}{10}
\providecommand{\url}[1]{#1}
\csname url@samestyle\endcsname
\providecommand{\newblock}{\relax}
\providecommand{\bibinfo}[2]{#2}
\providecommand{\BIBentrySTDinterwordspacing}{\spaceskip=0pt\relax}
\providecommand{\BIBentryALTinterwordstretchfactor}{4}
\providecommand{\BIBentryALTinterwordspacing}{\spaceskip=\fontdimen2\font plus
\BIBentryALTinterwordstretchfactor\fontdimen3\font minus
  \fontdimen4\font\relax}
\providecommand{\BIBforeignlanguage}[2]{{%
\expandafter\ifx\csname l@#1\endcsname\relax
\typeout{** WARNING: IEEEtran.bst: No hyphenation pattern has been}%
\typeout{** loaded for the language `#1'. Using the pattern for}%
\typeout{** the default language instead.}%
\else
\language=\csname l@#1\endcsname
\fi
#2}}
\providecommand{\BIBdecl}{\relax}
\BIBdecl

\bibitem{StHe:95}
S.~Staniford-Chen and L.~Heberlein, ``Holding intruders accountable on the
  internet,'' in \emph{Security and Privacy, 1995. Proceedings., 1995 IEEE
  Symposium on}, may 1995, pp. 39 --49.

\bibitem{LuZhZhPeLe:11}
X.~Luo, P.~Zhou, J.~Zhang, R.~Perdisci, W.~Lee, and R.~K.~C. Chang, ``Exposing
  invisible timing-based traffic watermarks with {BACKLIT},'' in
  \emph{Proceedings of the 27th Annual Computer Security Applications
  Conference}.\hskip 1em plus 0.5em minus 0.4em\relax ACM, 2011, pp. 197--206.

\bibitem{DoFlShPaCoStL:02}
D.~L. Donoho, A.~G. Flesia, U.~Shankar, V.~Paxson, J.~Coit, and S.~Staniford,
  ``Multiscale stepping-stone detection: detecting pairs of jittered
  interactive streams by exploiting maximum tolerable delay,'' in
  \emph{Proceedings of the 5th international conference on Recent advances in
  intrusion detection}, ser. RAID'02.\hskip 1em plus 0.5em minus 0.4em\relax
  Berlin, Heidelberg: Springer-Verlag, 2002, pp. 17--35.

\bibitem{BlSoVe:04}
A.~Blum, D.~Song, and S.~Venkataraman, ``Detection of interactive stepping
  stones: Algorithms and confidence bounds,'' in \emph{Recent Advances in
  Intrusion Detection}, ser. Lecture Notes in Computer Science.\hskip 1em plus
  0.5em minus 0.4em\relax Springer Berlin / Heidelberg, 2004, vol. 3224, pp.
  258--277.

\bibitem{MoOs:03}
P.~Moulin and J.~O'Sullivan, ``Information-theoretic analysis of information
  hiding,'' \emph{Information Theory, IEEE Transactions on}, vol.~49, no.~3,
  pp. 563--593, 2003.

\bibitem{BaTo:13}
M.~Barni and B.~Tondi, ``The source identification game: An
  information-theoretic perspective,'' \emph{IEEE Transactions on Information
  Forensics and Security}, vol.~8, no.~3, pp. 450--463, 2013.

\bibitem{BoAz:97}
A.~W. Bowman and A.~Azzalini, \emph{{Applied Smoothing Techniques for Data
  Analysis: The Kernel Approach with S-Plus Illustrations (Oxford Statistical
  Science Series)}}.\hskip 1em plus 0.5em minus 0.4em\relax Oxford University
  Press, USA, Nov. 1997.

\bibitem{KoHeAbYe:04}
D.~Kotz, T.~Henderson, I.~Abyzov, and J.~Yeo, ``{CRAWDAD} trace set
  dartmouth/campus/tcpdump (v. 2004-11-09),''
  \url{http://crawdad.cs.dartmouth.edu/dartmouth/campus/tcpdump}, Nov. 2004.

\bibitem{BaSaPrVa:10}
R.~R.~R. {Barbosa}, R.~{Sadre}, A.~{Pras}, and R.~{van de Meent},
  ``Simpleweb/university of twente traffic traces data repository,'' Centre for
  Telematics and Information Technology University of Twente, Enschede,
  Technical Report, 2010.

\bibitem{ElPe:13}
J.~A. Elices and F.~P\'erez-Gonz\'alez, ``Measures to model delays on
  internet,'' \url{http://www.unm.edu/~elices/captures.html}, Jan. 2013.

\end{thebibliography}
\end{spacing}

\end{document}